\documentstyle[12pt,epsfig,axodraw,lineno]{article}
\advance\textheight by \topskip
\begin{document}
%\linenumbers
\tolerance=100000
\thispagestyle{empty}
\setcounter{page}{0}

\def\cO#1{{\cal{O}}\left(#1\right)}
\newcommand{\be}{\begin{equation}}
\newcommand{\ee}{\end{equation}}
\newcommand{\br}{\begin{eqnarray}}
\newcommand{\er}{\end{eqnarray}}
\newcommand{\ba}{\begin{array}}
\newcommand{\ea}{\end{array}}
\newcommand{\bi}{\begin{itemize}}
\newcommand{\ei}{\end{itemize}}
\newcommand{\bn}{\begin{enumerate}}
\newcommand{\en}{\end{enumerate}}
\newcommand{\bc}{\begin{center}}
\newcommand{\ec}{\end{center}}
\newcommand{\ul}{\underline}
\newcommand{\ol}{\overline}
\newcommand{\ra}{\rightarrow}
\newcommand{\sm}{${\cal {SM}}$}
\newcommand{\as}{\alpha_s}
\newcommand{\aem}{\alpha_{em}}
\newcommand{\ycut}{y_{\mathrm{cut}}}
\newcommand{\susy}{{{SUSY}}}
\newcommand{\Dir}{\kern -6.4pt\Big{/}}
\newcommand{\Dirin}{\kern -10.4pt\Big{/}\kern 4.4pt}
\newcommand{\DDir}{\kern -10.6pt\Big{/}}
\newcommand{\DGir}{\kern -6.0pt\Big{/}}
\def\Ecm{\ifmmode{E_{\mathrm{cm}}}\else{$E_{\mathrm{cm}}$}\fi}
\def\gluino{\ifmmode{\mathaccent"7E g}\else{$\mathaccent"7E g$}\fi}
\def\photino{\ifmmode{\mathaccent"7E \gamma}\else{$\mathaccent"7E \gamma$}\fi}
\def\gl{\ifmmode{m_{\mathaccent"7E g}}
             \else{$m_{\mathaccent"7E g}$}\fi}
\def\taugluino{\ifmmode{\tau_{\mathaccent"7E g}}
             \else{$\tau_{\mathaccent"7E g}$}\fi}
\def\mphotino{\ifmmode{m_{\mathaccent"7E \gamma}}
             \else{$m_{\mathaccent"7E \gamma}$}\fi}
\def\ML{\ifmmode{{\mathaccent"7E M}_L}

             \else{${\mathaccent"7E M}_L$}\fi}
\def\MR{\ifmmode{{\mathaccent"7E M}_R}
             \else{${\mathaccent"7E M}_R$}\fi}
\def\lsim{\buildrel{\scriptscriptstyle <}\over{\scriptscriptstyle\sim}}
\def\gsim{\buildrel{\scriptscriptstyle >}\over{\scriptscriptstyle\sim}}
\def\jp #1 #2 #3 {{J.~Phys.} {#1} (#2) #3}
\def\pl #1 #2 #3 {{Phys.~Lett.} {#1} (#2) #3}
\def\np #1 #2 #3 {{Nucl.~Phys.} {#1} (#2) #3}
\def\zp #1 #2 #3 {{Z.~Phys.} {#1} (#2) #3}
\def\pr #1 #2 #3 {{Phys.~Rev.} {#1} (#2) #3}
\def\prep #1 #2 #3 {{Phys.~Rep.} {#1} (#2) #3}
\def\prl #1 #2 #3 {{Phys.~Rev.~Lett.} {#1} (#2) #3}
\def\mpl #1 #2 #3 {{Mod.~Phys.~Lett.} {#1} (#2) #3}
\def\rmp #1 #2 #3 {{Rev. Mod. Phys.} {#1} (#2) #3}
\def\sjnp #1 #2 #3 {{Sov. J. Nucl. Phys.} {#1} (#2) #3}
\def\cpc #1 #2 #3 {{Comp. Phys. Comm.} {#1} (#2) #3}
\def\xx #1 #2 #3 {{#1}, (#2) #3}
\def\NP(#1,#2,#3){Nucl.\ Phys.\ \issue(#1,#2,#3)}
\def\PL(#1,#2,#3){Phys.\ Lett.\ \issue(#1,#2,#3)}
\def\PRD(#1,#2,#3){Phys.\ Rev.\ D \issue(#1,#2,#3)}
\def\preprint{{preprint}}
\def\Ord{\lower .7ex\hbox{$\;\stackrel{\textstyle <}{\sim}\;$}}
\def\OOrd{\lower .7ex\hbox{$\;\stackrel{\textstyle >}{\sim}\;$}}
%%% PArticle Definition
\def \MCH {$\tilde\chi_1^+$}
\def \CH{{\tilde\chi}^{\pm}}
\def \N0{\tilde\chi^0}
\def \LSP{\tilde\chi_1^0}
\def \SNU{\tilde{\nu}}
\def \BARSNU{\tilde{\bar{\nu}}}
\def \MLSP{m_{{\tilde\chi_1}^0}}
\def \MCH{m_{{\tilde\chi}^{\pm}}}
\def \MCHMIN {\MCH^{min}}
\def \ET{\not\!\!{E_T}}
\def \LL{\tilde{l}_L}
\def \LR{\tilde{l}_R}
\def \MLL{m_{\tilde{l}_L}}
\def \MLR{m_{\tilde{l}_R}}
\def \MSNU{m_{\tilde{\nu}}}
\def \PROCESS{e^+e^- \rightarrow \tilde{\chi}^+ \tilde{\chi}^- \gamma}
\def \PI{{\pi^{\pm}}}
\def \DM{{\Delta{m}}}
\newcommand{\bQ}{\overline{Q}}
\newcommand{\ad}{\dot{\alpha }}
\newcommand{\bd}{\dot{\beta }}
\newcommand{\dd}{\dot{\delta }}
\def \sq{\tilde q}
\def \gl{\tilde g}
\def \LSP{\tilde\chi_1^0}
\def \MUL{m_{\tilde{u}_L}}
\def \MUR{m_{\tilde{u}_R}}
\def \MDL{m_{\tilde{d}_L}}
\def \MDR{m_{\tilde{d}_R}}
\def \MSNU{m_{\tilde{\nu}}}
\def \MTAUL{m_{\tilde{\tau}_L}}
\def \MTAUR{m_{\tilde{\tau}_R}}
\def \mhf{m_{1/2}}
\def \MST{m_{\tilde t_1}}
\def \CHM{H^\pm}
\def \RPVC{\lambda'}
\def\tth{\tilde{t}\tilde{t}h}
\def\qqh{\tilde{q}_i \tilde{q}_i h}
\def\t1{\tilde t_1}
\def \ta1{\tilde\tau_1}
\def \MET{E{\!\!\!/}_T}  
\def \invfb{fb^{-1}}
\def\bul{\bullet}
%#############################
%#$\tan\beta \approx 3$ 
\def\lapp{\mathrel{\rlap{\raise.5ex\hbox{$<$}}
                    {\lower.5ex\hbox{$\sim$}}}}
\def\gapp{\mathrel{\rlap{\raise.5ex\hbox{$>$}}
                    {\lower.5ex\hbox{$\sim$}}}}
%#############################
\begin{flushright}
%\today
\end{flushright}
\begin{center}
{\Large \bf
Probing a Mixed Neutralino Dark Matter Model at the 7 TeV LHC
} 
\\[1.00cm]
\end{center}
\begin{center}
{\large Monoranjan Guchait$^a$,D.P. Roy$^b$, Dipan Sengupta$^a$  
}\\[0.3 cm]
{\it 
\vspace{0.2cm}
$^a$ Department of High Energy Physics\\
Tata Institute of Fundamental Research\\ 
Homi Bhabha Road, Mumbai-400005, India.\\
\vspace{0.2cm}
$^b$ Homi Bhabha's Centre for Science Education \\
Tata Institute of Fundamental Research \\
V. N. Purav Marg, Mumbai-400088, India.
}
\end{center}

\vspace{2.cm}

\begin{abstract}
{\noindent\normalsize 
We have analyzed the prospect of probing a non-universal gaugino mass model
of mixed bino-higgsino dark matter at the current 7 TeV run of LHC. It 
provides cosmologically compatible dark matter relic density over two 
broad bands of parameters, corresponding to $m_{\gl} < m_{\sq}$ and 
$m_{\gl} \sim m_{\sq}$. The SUSY spectrum of this model has two distinctive 
features : (i) an approximate degeneracy among the lighter chargino and 
neutralino masses, and (ii) an inverted mass hierarchy of squark masses. 
We find that these features can be exploited to obtain a viable signal 
upto $m_{\gl} \sim$ 800 GeV over both the parameter bands with an 
integrated luminosity 5$\invfb$. 
}
\end{abstract}
\vspace{2cm}
\hskip1.0cm
%PACS no: 11.30.pb, 14.60.Cd, 14.80.Ly
%\vspace*{\fill}
%\vskip1.0cmmey
%\noindent
%\vspace*{\fill}
\newpage
\section{Introduction}
\label{sec_intro}
Supersymmetry is a very popular and well motivated extension
of the Standard Model(SM) as it provides a natural solution to
the gauge hierarchy problem of the SM along with
a natural candidate for the dark matter(DM) of the universe.
This has led to an intense global search for supersymmetry(SUSY) 
in the colliders as well as the dark matter experiments,
which are both going through rapid developments. The CERN Large 
Hadron  Collider(LHC) has already accumulated a luminosity of
~3 $fb^{-1}$ at a CM energy of 7 TeV, which is expected to go up to
$5 fb^{-1}$ by the end of this year and 10-15$\invfb$ 
by the end of the current run in late 2012. After this  the 
machine is scheduled to be upgraded to run at the designed CM energy of 14 
TeV starting in 2015. Then one can probe most of the interesting parameter 
space of supersymmetry and in particular the minimal supersymmetric standard 
model(MSSM). Likewise direct DM detection experiments like XENON100 and Super 
CDMS as well as the indirect detection experiment like Ice Cube and Fermi-LAT 
are expected to probe significant parts of the MSSM parameter space over 
the next 4-5 years. It is well known now that the so called constrained MSSM 
or the minimal supergravity(mSUGRA) model corresponds to a dominantly bino 
DM, resulting in a generic overabundance of   
relic density. There are only a few narrow strips of the mSUGRA 
parameter space that correspond to cosmologically compatible DM relic 
density. It will be possible to probe these cosmologically compatible regions 
of mSUGRA parameter space at the 14 TeV runs of LHC, but not at the 7 TeV. 
In fact the 7 TeV run of LHC is expected to probe only a limited region of the 
mSUGRA parameter space, much of which is already excluded by the 
Higgs mass limit from LEP~\cite{tata,lhc}. 
\\

Moreover, much of these cosmologically compatible parameter regions of 
mSUGRA 
model will be hard to probe at the above mentioned DM experiments. On
 the other hand there are simple and well motivated versions of MSSM with  
non-universal gaugino masses at the GUT scale, which correspond to mixed 
bino-higgsino dark matter. They lead to cosmologically  
compatible DM relic density over large parts of parameter space and also 
promising signals in the above mentioned DM experiments. In this paper we 
shall study the prospect of probing such a mixed bino-higgsino dark matter 
model at the current 7 TeV run of LHC. We find a promising signal 
for this model 
over a significant part of the cosmologically compatible parameter space. 
\\

In section 2 we summarize the essential features of the two non-niversal 
gaugino mass models with mixed neutralino dark matter, which give 
cosmologically compatible relic density along with promising direct detection 
signals over large parts of their parameter space.  We briefly discuss these 
features and for one of them, which corresponds to a relatively light 
gluino and hence a 
promising signal at 7 TeV. In section 3 we study in detail the signal cross 
sections along with the relevant SM backgrounds 
in jets + missing transverse energy($\MET$) as well as leptonic 
channels. We conclude 
with a brief summary of our results in section 4.  
\section{Non-universal Gaugino mass Models for Mixed Neutralino DM}
These models are based on the assumption that SUSY is broken by an admixture 
of two superfields belonging to a singlet and a non-singlet representation of 
the GUT group which is SU(5) for the simplest case~\cite{ellis}. 
The gauge kinetic 
function responsible for the GUT scale gaugino masses originates from the 
vacuum expectation value of the F term of a chiral superfield $\Phi$ 
responsible for SUSY breaking,
\br
 \frac{\langle F_{\Phi} \rangle}{M_{Planck}} \lambda_{i}\lambda_{j}
\er
 where $\lambda_{1,2,3}$ are the U(1), SU(2), SU(3) gaugino fields - bino, 
wino and gluino. Since the gauginos belong to the adjoint representation 
of the GUT group SU(5), $\Phi$ and $F_{\Phi}$ can belong to any of the 
irreducible 
representations appearing in their symmetric product.
\begin{equation}
(24\times24)_{sym}=1+24+75+200.  
\end{equation}
Thus the GUT scale gaugino masses for a given representation are determined 
in terms of a given SUSY breaking mass parameter by
\br
M^{G}_{1,2,3}=C^{n}_{1,2,3}m^{n}_{1/2}
\label{eq:gino}
\er
where,
\br
C^{1}_{1,2,3}=(1,1,1); C^{24}_{1,2,3}=(-1,-3,2);C^{75}_{1,2,3}=(-5,3,1);
 C^{200}_{1,2,3}=(10,2,1).
\label{eq:cs}
\er

The mSUGRA model assumes $\Phi$ to be a singlet, leading to universal 
gaugino masses at the GUT scale. On the other hand any of the non-singlet 
representations for $\Phi$ would imply non-universal gaugino masses as per 
eqs.~\ref{eq:gino} and \ref{eq:cs}. The phenomenology of such non-universal 
gaugino mass models 
have been extensively studied in the 
literature~\cite{nugm,utpal,king,das,roy}. 
Since the gaugino 
masses evolve like the corresponding gauge couplings at the one-loop level  
of the RGE, the three gaugino masses are proportional to the respective gauge 
couplings, i.e,
\br
M_{1} &=& (\alpha_{1}/\alpha_{G})M_1^G \simeq (25/60)C_{1}^{n}m_{1/2}^{n},
\nonumber \\
M_{2} &=&  (\alpha_{2}/\alpha_{G})M_2^G \simeq (25/30) C_{2}^{n}m_{1/2}^{n},
\nonumber \\
M_{3} &=&  (\alpha_{3}/\alpha_{G})M_3^G \simeq (25/9) C_{3}^{n}m_{1/2}^{n}.
\label{eq:Ms}
\er 

The higgsino mass parameter $ \mu $ is obtained from the electro-weak(EW) 
symmetry breaking 
condition along with the one loop RGE for the Higgs scalar mass, i.e,
\br
\mu^{2}+M_{Z}^{2}/2 \simeq -m_{H_{u}}^{2} \simeq -0.1m_{0}^{2}+2.1M_{3}^{G^{2}}
 -0.22M_{2}^{G^{2}} + 0.19M_{2}^{G}M_{3}^{G}
\label{eq:ewsb}
\er
neglecting the contribution from the GUT scale trilinear coupling 
term $A_{0}$~\cite{carena}. The numerical co-efficients on the 
right correspond to a representative value of $\tan\beta = 10$; but they 
show only mild variation 
over the moderate $\tan\beta$ region.
\\

Although we use exact numerical solutions to the two-loop RGE in our 
analysis, the composition of the lightest neutralino DM $\chi_{1}^{0}$ 
can be seen from the relative values of the gaugino and higgsino masses,
given by eqs.~\ref{eq:gino}-\ref{eq:Ms} and eq.\ref{eq:ewsb} 
respectively. For the mSUGRA(singlet $\Phi$), 
one gets  $M_{1} < \mu$, resulting in a bino DM over most of the parameter 
space. Since bino has no gauge charge it can only pair annihilate via  
sfermion exchange and the large sfermion mass limit from 
LEP~\cite{amsller} 
leads to overabundance of DM relic density. The same is true for the 
24-plet representation. On the other hand, the 75 and 200 plet 
representations imply $M_{1,2} > \mu$, leading to higgsino DM over 
most of the parameter space, and since the higgsino DM can effectively 
co-annihilate with its nearly degenerate chargino via W-boson exchange, 
one gets a generic underabundance of DM relic density for these two 
representations~\cite{utpal}.
Finally, assuming the SUSY breaking to occur via an admixture of a 
singlet and a non-singlet superfield belonging to the (1+75) 
or (1+200) representations~\cite{king,das}, i.e,
\br
m_{1/2}^{1}&=&(1-\alpha_{75})m_{1/2};\ \  m_{1/2}^{75}=\alpha_{75}m_{1/2},
\\ \nonumber
m_{1/2}^{1}&=&(1-\alpha_{200})m_{1/2};\ \  
m_{1/2}^{200}=\alpha_{200}m_{1/2},      
\label{eq:mhalf}
\er 
one can get large admixture of gaugino and higgsino components in the DM. 
It was shown in~\cite{das} that one gets optimal admixture of bino and 
higgsino 
components in the (1+75) model and bino, wino, higgsino components in the 
(1+200) model with $\alpha_{75}=0.475$ and $\alpha_{200}=0.12$, leading 
to WMAP~\cite{wmap} satisfying DM relic density over large parts of the MSSM 
parameter space. They are two simple realizations of the so called well 
tempered neutralino scenario\cite{arkani}. Once the mixing parameter is 
fixed, each of these models is as predictive as mSUGRA model. The WMAP 
satisfying DM 
relic density region of the (1+200) model correspond to relatively 
heavy gluino and 
squark masses in the TeV range, which is inaccessible at the 7 TeV run. We 
therefore concentrate on the (1+75) model. 
\\

The large admixture of the bino and the higgsino components of the DM in 
this model leads to promising signals for the direct detection experiments as 
well as indirect detection experiments like Ice Cube~\cite{das}. In 
particular, the model prediction was shown to agree 
with the putative direct DM signal, corresponding to the two CDMS II 
candidate events~\cite{cdms}, over the relatively low DM  mass 
region~\cite{roy}. Figure 1 compares the model 
prediction with the putative signal corridor, represented by the central 
value and the 90$\%$ upper and lower limits corresponding to the two 
CDMS II candidate events. The 90\% upper limit of the more recent 
XENON100 experiments~\cite{xenon} is overlayed on this figure for 
comparison. One sees that the central value curve of the two CDMS II 
candidate events is in conflict 
with the XENON100 upper limit, which lies marginally below it. However there 
are only very few points of the model parameter scan that overlap with this 
central value. Most of the  points are crowded near the lower limit of the 
putative signal corridor of the two CDMS II candidate events, which are 
also compatible with the XENON100 upper limit. One expects this part of the 
model prediction to be tested by the XENON100 and the super CDMS 
experiments over the next few years. It is therefore very pertinent 
to find out if this part of the model parameter space can be probed at 
the current LHC run. 
\begin{figure*}[t]
\centering
\includegraphics[height=3.0in,width=4.0in]{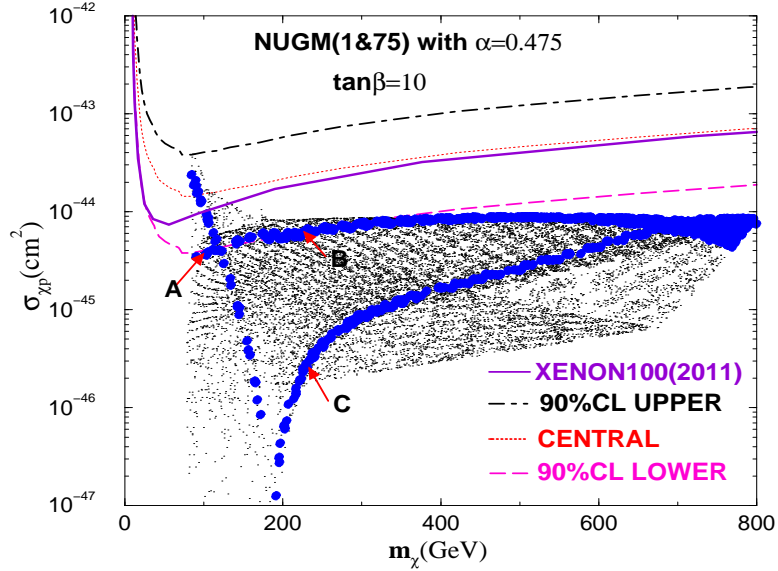}
\caption{
Prediction of the (1+75) model compared with the putative signal corridor 
corresponding to the two candidate DM scattering events of the CDMS II 
experimental~\cite{cdms}.  The 90\% C.L. upper limit of the recent XENON100
experiments~\cite{xenon} is overlayed as the solid line. 
The blue(dark) dots corresponding the WMAP relic density satisfying 
region of the model.
The three representative points A,B and C are shown in the low mass part of
the region which is accessible to the 7 TeV LHC run.}
\label{fig:fig1}
\end{figure*}
It should be noted here that the upper band of the WMAP relic density 
compatible points of the (1+75) model shown in figure 1 corresponds 
to a DM with comparable bino and higgsino components, which 
annihilate via(s and t channel) gauge boson exchange processes
\begin{equation}
\chi_{1}^{0}\chi_{1}^{0},\chi_{1}^{0}\chi_{2}^{0},\chi_{1}^{0}\chi_{1}^{\pm}
\rightarrow WW,ZZ,f\bar{f}.
\end{equation}
On the other hand, the lower band of the WMAP relic density compatible
points corresponds to a DM with $\le 10 \%$ higgsino component, which 
annihilate via s channel Higgs exchange processes[8] 
\br
\chi_{1}^{0}\chi_{1}^{0},\chi_{1}^{0}\chi_{2}^{0},\chi_{1}^{0}\chi_{1}^{\pm}
\rightarrow t\bar{t},b\bar{b},\tau\bar{\tau},t\bar{b}, \tau\nu_{\tau}.
\er
The comparable gaugino and higgsino components of the DM for the upper 
branch accounts for the large spin independent $\chi p$ scattering cross 
section, while the low higgsino component of DM in the lower branch 
accounts for the low $\chi p$ cross section. Therefore, it will be more 
difficult to probe the parameter space corresponding to lower branch in 
direct DM detection experiments. On the other hand both the branches can 
be probed at the 7 TeV LHC run up to a gluino mass of $\sim 800$ GeV 
corresponding to $m_{\chi} \sim$ 250~GeV. In fact the lower branch will 
be easier to probe since the corresponding cross section is larger due to 
the lower squark masses($m_{\sq} \sim m_{\gl}$) as we shall see below. 
Therefore we have selected two representative points 
on the upper branch along with one on the lower branch spanning the 
above mentioned gluino mass range. In the next section we shall 
investigate in detail the 7 TeV LHC 
signals corresponding to these three representative points in the parameter 
space of interest along with relevant backgrounds.
\section{The expected model signals and backgrounds in the 7 TeV LHC run}
%Event Simulation, Signal and Background }
In table 1, we show the mass spectrum of SUSY particles at the electro-weak
scale for the three representative points A, B and C marked in figure 1.
Table 2 shows the corresponding decay branching ratios. We use the 
{\tt SuSpect-SUSYHIT}~\cite{suspect} to generate the SUSY 
spectra and branching ratios.
We see from table 1 that the SUSY spectrum of the (1+75) model has two 
distinctive features vis a vis the mSUGRA model -(1) a near degeneracy of 
the lighter neutralino and chargino($\chi_{1}^{0},\chi_{2}^{0},\chi_{1}^{\pm}$)
masses and (2) inverted hierarchy of the squark masses where in particular 
$\tilde{t_{1}}$ is significantly lighter than the other squarks. Indeed both 
the features are common to most mixed neutralino DM models. The 1st feature 
implies that $\chi_{1}^{0}$ coming from the SUSY cascade decay carries a 
relatively a large fraction of the energy-momentum, resulting in a hard 
missing transverse energy ($\MET$) distribution. The 2nd feature implies 
that the gluino decays preferentially to t and b 
quarks leading to  several hard isolated leptons and b-tags along with 
the large $\MET$. These features make the LHC signal for the mixed 
neutralino DM model more promising in comparison to the much studied 
mSUGRA  model. 
% table 1
\begin{table}
\begin{center}
\begin{tabular}{|c|c|c|c|c|c|c|c|c|c|c|c|}
\hline
Model & $\tilde{g}$ & $\tilde{q}_L$ &$\tilde q_R$  & $\tilde{t}_1$ & 
$\tilde{b}_1$ 
& $\tilde{e}_l$ & $\tilde{\tau_{1}}$ & 
${\chi}_{1}^{0}$&${\chi}_{2}^{0}$
& ${\chi}_{1}^{+} $& ${\chi}_{2}^{+}$ \\
\hline
A & 433 & 1280&1274 & 759  & 1054 & 1263 & 1246 &104 &122
&123 & 271  \\
\hline
B &793 &1480&1440 &902 & 1246 & 1375 & 1327 & 227 &256 & 257 &501 \\
\hline
C & 722 & 750 &660 & 483 & 649 & 437 & 237 & 231 & 301 & 302 & 490 \\
\hline  
\end{tabular}
\caption{ 
Mass spectrum(in GeV) for the three representative parameter points 
(A) $m_{1/2}=$144~GeV,
$m_0$=1255 GeV, (B) $m_{1/2}=300$~GeV, $m_0=$1325~GeV, 
(C) $m_{1/2}=300$~GeV, $m_{0}=$185~GeV.
}
\end{center}
\end{table}
Table 2 shows that point A has a large two body decay
$\tilde{g}\rightarrow g\chi_{2,3}^{0}$. This arises via the the 
virtual $\tilde{t}\bar{t}$ channel since the $t\bar{t}\chi_{1}^{0}$ 
channel is kinetically forbidden in this case. However the decay channels
\begin{equation}
\tilde{g}\rightarrow \bar{t}t\chi_{i}^{0}, \bar{t}b(t\bar{b})\chi_{1}^\pm,
b\bar{b}\chi^{0}_{2}  
\end{equation}
\begin{table}
\begin{center}
\begin{tabular}{|c|c|c|c||}
\hline
Decays & A & B & C\\
\hline
$ \tilde{g} \rightarrow \tilde{q} {q}$& & &$43\%$ \\
$ \tilde{g} \rightarrow \tilde{b}{b} $& & & $17\%$\\

$ \tilde{g} \rightarrow \tilde{t} {t} $& & & $40\%$\\

$ \tilde{t_{1}} \rightarrow b\chi^{\pm}_{1} $& & & $88\%$\\

$ \tilde{q_{L}} \rightarrow q \chi^{\pm}_{1,2} $& & & $62\%$\\

$ \tilde{q_{L}} \rightarrow q\chi^{0}_{1,2,3,4} $& & & $32\%$\\

$ \tilde{g} \rightarrow t\bar{b}\chi_{1}^{\pm} $ & $ 25\% $& $54\% $ &\\

$ \tilde{g}\rightarrow g\chi_{2,3}^{0} $ & $ 30 \% $ & $4.2\%$ &  \\
$\tilde{g}\rightarrow q\bar{q}\chi_{1,2}^{\pm} $ & $16\% $ & $2.6\%$&\\
$\tilde{g}\rightarrow b\bar{b}\chi_{1,2,3,4}^{0} $ & $12\% $& $2\%$ & \\
$\tilde{g}\rightarrow t\bar{t}\chi_{1,2,3,4}^{0} $ & & $27\%$ & \\
$\tilde{g}\rightarrow q\bar{q}\chi_{1,2,3,4}^{0} $ & $17\% $&$2\%$ & \\
$\chi_{1}^{+} \rightarrow q\bar{q}\chi_{1}^{0} $ & $68 \%$ &$67\%$&\\
$\chi_{1}^{+} \rightarrow e\bar{e}\chi_{1}^{0} $ & $22\%$ & $22\%$&\\
$\chi_{1}^{+} \rightarrow \tilde{\tau}\nu_{\tau} $ & &  &$100\%$\\
$\chi_{2}^{0} \rightarrow q\bar{q}\chi_{1}^{0} $ & $ 68 \% $&$66\%$ &\\
$ \chi_{2}^{0}\rightarrow \tilde{\tau_{1}}\tau $ & & & $ 84\%$ \\
$ \chi_{2}^{0}\rightarrow \tilde{e}e $ & & & $ 16\%$ \\
\hline
\end{tabular}
\caption{Decay branching ratios of sparticles 
for three representative points A, B and C, as shown in table 1.}
\end{center}
\end{table}
account for $83 \%$ for point B and $57 \%$ for point C. Therefore one 
can effectively enhance the $\tilde{g}\tilde{g}$ pair production signal 
over background by demanding $\ge$3 b-tags. As we shall see below, this 
is particularly useful for the point B where the SUSY signal is dominated 
by $\tilde g\tilde{g}$ pair production.
The pair production cross sections of strong SUSY particles,
$\gl\gl, \sq\gl, \sq\sq,\sq\sq^\ast$ are calculated using
{\tt Prospino}~\cite{prospino} at the NLO level.   
\\

The signal and background events are simulated using {\tt PYTHIA}~\cite{pythia}
interfacing with fast detector simulation package {\tt PGS}~\cite{pgs}. In 
{\tt PGS} 
jets are reconstructed using cone algorithm with cone size 
$\Delta R=$0.5, taking inputs from the energy deposits in the 
calorimeter cells. In the simulation jets are selected with
\br 
p_T^j \ge 30{\rm ~GeV};~|\eta^j|\le 3.0.
\label{eq:jcut}
\er
We select leptons($e$ and  $\mu$) with 
\br
p_{T}^\ell \ge 10{\rm~GeV}; ~|\eta^\ell| \le 2.5,
\label{eq:lcut}
\er
and for di-lepton final state leading lepton to pass a cut  
\br 
p_{T}^{\ell_1} \ge 20{\rm~GeV}.
\label{eq:llcut}
\er
The isolation of lepton is ensured by looking at the total transverse 
energy $E_{T}^{ac}\le 20 \%$ of the $p_T$ of lepton,
where $E_{T}^{ac}$ is the scalar sum of transverse energies of jets 
within a cone of size $\Delta R(l,j) \le 0.2 $ between jet and lepton.
In PGS the missing transverse energy($\MET$) is measured 
using the toy calorimeter informations. While the use of exact  
b-tagging method is out of the scope of this work, 
the b jet candidate is identified by performing a matching 
between a b quark and jet with a matching cone $\Delta R=$0.3.
The selected 3 and 4 b jets events are multiplied by efficiency factors
$\epsilon_b^3$=0.21 and $4{\epsilon_b}^3 - 3{\epsilon_b}^4$=0.47 
for $\ge$3 b-tags assuming
b tagging efficiency $\epsilon_b$=0.6~\cite{btag}.
The dominant SM backgrounds corresponding to signal 
originate from the $t \bar t$ and QCD processes. We have checked also
other SM processes like, W/Z+jets, WW,WZ,ZZ, whose contribution to
the total background cross section is found to be negligible after applying 
selection cuts.      
We investigate the discovery potential of the signal for 
this mixed neutralino DM model in 
the dii-lepton and single lepton channels including jets plus $\MET$.
We also analyze the inclusive jets+$\MET$ signal which is expected 
to give better sensitivity. As explained above, 
we require $\ge$ 3 b-tagged jets for single lepton and jets+$\MET$ 
final states to suppress SM backgrounds mainly from 
$t \bar t$.
In the following we describe 
search strategies for all the three channels.
\\

{\bf Di-lepton + jets + $\MET$:}
In the di-lepton channel we select both 
same sign(SS) and opposite sign(OS) leptons passed by   
lepton cuts, eqs.~\ref{eq:lcut},~\ref{eq:llcut}. 
In addition, these events are also subject to following 
kinematic cuts to suppress SM backgrounds,
\br
{\rm Number~of~jets} \ge 3;  \MET \ge 100 {\rm~GeV}, \nonumber \\ 
M_{eff} \ge 500 {\rm~GeV}; R_{\ell\ell} \le 0.25, \nonumber \\
m_{\ell\ell} \ne 75-110~{\rm~GeV~for~OS~leptons},
\label{eq:dilepcuts}
\er
where effective mass of the event 
$M_{eff} = \sum_{j=1}^{3} p_T^j + E{\!\!\!/}_T$  and  
$R_{\ell\ell}=\frac{P_T^{\ell_1} + P_T^{\ell_2}}{H_T}$ with 
$H_T = \sum_{1}^{n_j} p_{T}^{j_i}$.
In signal events
multiplicity of hard jets are expected to be higher 
than the SM processes, so $R_{\ell\ell}$ is likely to be distributed towards
the lower values which is reflected in the suppression of $t \bar t$ background
due to this $R_{\ell\ell}$ cut. 
An invariant mass($m_{\ell\ell}$) cut for OS lepton case is applied to 
suppress the background 
from Z decay. In table 3 we present the effects of 
each set of cuts described by eq.\ref{eq:dilepcuts} to the signal 
process for the parameter points A,B and C and as well as the 
backgrounds. 
% table 3
\begin{table}
\begin{center}
\begin{tabular}{|c|c|c|c|c|c|c|c|c|c|c|c|}
\hline
Proc& C.S & $N_{ev}$ & 2$\ell$ \&  & $\MET\ge$ &Meff &
$R_{ll}\le$ &SS,OS &$m_{\ell\ell}$ & 1$\invfb$ & 5$\invfb$ \\
&  (pb) & &$n_j \ge3 $ & 100 & $\ge$500 &
0.25 &  &$\ne$ 75-110 & SS~OS &SS~OS \\
\hline

A:$\tilde{g}\tilde{g}$ & 5.84 & 10k &185 &134 &75 &61 &16,45 &16,36 &
10.0,21.0 &50.,105 \\  
A:$\tilde{q}\tilde{g}$ & 0.28 & 10k & 342 & 303 & 301 &279 
&89,190&89,150 &2.5, 4.2&12.5,23 \\
\hline
Total & & & & & & & & &12.5,25.2 & 62.5,126.0  \\
\hline

B:$\tilde{g}\tilde{g}$ & 0.06 & 10k &737 &609 &503 &382 &142,240 
&142,198 &1.0,1.3 &5,6.5 \\
B:$\tilde{q}\tilde{g}$ & 0.018 & 10k &492 &359 &317 &290 &12,188 
&12,111 & 0,.2 &0.,1.0 \\
\hline
Total & & & & & & & & & 1.0, 1.5 & 5.,7.5 \\  

\hline
C:$\tilde{g}\tilde{g}$ & 0.1 & 10k &401 &351 &306 &233 &41,192 
&41,135 & 0.6,1.5 &3.0,7.5 \\
C:$\tilde{q}\tilde{g}$ & 0.57 & 10k &492 &359 &317 &290 &12,188 
&12,111 &
 0.75,6.3 & 3.75,31.5 \\
C:$\tilde{q}\tilde{q}$ & 0.55 & 10k &395 &348 &333 &234 &13,221 
&13,145 & 0.8,8 & 4.,40. \\
\hline
Total & & & & & & & & & 2.15,15.8 &10.75,79 \\
\hline  
$t\bar{t}$ & & &  & & & & & & & \\
\hline
$5-200$ & 143.5 & 100k & 227 &78&31&13&2,11&2,9& 6.0,21.0 & 30.,105  \\
200-500 & 16.3 & 50k & 292 & 145&87&33&5,28&5,22 & 2.5, 10.7 & 12.5,53.5 \\
500-inf & 0.16 & 10k  & 87 & 65 &63 &26 & 6,20&6,19& 0.5,1.3 & 2.5, 6.5\\
\hline
Total & & & & & & & & & 9.0,33.0 & 45.0,165.0\\
\hline
QCD &  & & & & & & & & & \\
\hline
200-300 &6983& 1M   &1 &1&0&0&0,0&0,0&0,0 &0.,0. \\
300-500&873&1M &15 &2 &2 &1 &1,0&1,0&0.9,0 &4.5.,0. \\
500-800&43.1&100k &4 &2 &2 &0 &0,0&0,0&0,0 &0.,0. \\
\hline
Total &  & & & & & & & &0.9,0.0 & 4.5,0. \\
\hline
\end{tabular}
\caption{
The signal and SM background events for di-lepton final states. 
The $t \bar t$ and QCD backgrounds events are 
simulated for different $\hat p_T$ bins as shown. 
In the case of $t\bar t$ background, we multiply cross sections
by the K-factor 1.6 to take into account NLO effects~\cite{kidonakis}.
(Energy units are in GeV).}
\end{center}
\end{table}
The second and third columns present the production cross sections and
number of events simulated in each case respectively and the subsequent
rows show the suppression due to cuts. Clearly, the new cut $R_{\ell\ell}$
is killing backgrounds by a factor of two or more with a less effect
on the signal events. We present the survived number of events with
OS and SS sign leptons separately. 
Eventually, the last two
columns display the predicted number of signal and background events for
luminosity ${\cal L}=$1$\invfb$ and 5$\invfb$ respectively. 
They are normalized to the cross-sections of second column.
The last two columns show promising SS and OS di-lepton signals for point
A, with a signal/background ratio $>$1 for the SS channel. They also show 
a moderate OS di-lepton signal for the point C at 3(5)$\sigma$ level
with 1(5)$\invfb$ luminosity data. There is no viable di-lepton signal 
for the point B.   
\\

{\bf Single lepton + jets + $\MET$:}
We simulate the signal events in the single lepton final states along 
with jets and $\MET$. In addition to the selection of single 
lepton with cuts of eq.~\ref{eq:lcut}, 
we apply the following cuts,
\br
{\rm Number ~of ~jets }\ge 4 {\rm ~with} ~p_T^{j_1} \ge 100{\rm ~GeV},
\nonumber \\
m_T > 70{\rm ~GeV}, \MET > 150{\rm ~GeV}, M_{eff} \ge 500{\rm ~GeV}, 
\nonumber \\
R_{\ell} < 0.1 {\rm ~and} ~{\ge \rm ~3 ~ b-tagged ~jets},
\er
where the transverse mass between lepton and $\MET$ is defined as
$m_T = \sqrt{2 E_T^{\ell} \MET (1 - \cos\phi(\ell,\MET}))$ and
$R_{\ell}=\frac{P_T^{\ell_1}}{H_T}$. In table 4 we show the event 
summary for this case like table 3. The last two columns show
the number of events for the two values of luminosities and 
with $\ge$ 3 b-tagged jets in the final state. 
We have checked that without the $\ge$ 3 b-tags the signal and 
background sizes 
are very similar to the OS di-lepton case. The $\ge$ 3 b-tags effectively 
suppresses the background to $\le$50\% of the signal, so that the 
viability of the signal is determined by the number of signal events.
With 5$\invfb$ luminosity one expects $\sim$ 10 signal events for the point
A but only $\sim$ 2 events for points B and C. With a luminosity of 
15$\invfb$ expected in 2012 one would expect a viable signal of 5-6 
events for the points B and C.     
% Table 4
\begin{table}
\begin{center}
\begin{tabular}{|c|c|c|c|c|c|c|c|c|c|c|c|}
\hline
Proc&  C.S & N & Single & $n_{j}$  & $m_{T}$& $ \MET $&$M_{eff}$ &$R_{l}$&$n
_{b}$ & 1$\invfb$ & 5 $\invfb$   \\
&(pb) & &lep. &$\ge$4  & $\ge70$&$\ge150$
&$\ge500$    &$\le0.1$   &$\ge$ 3 &  &     \\
\hline
A:$\tilde{g}\tilde{g}$ & 5.84 & 10k & 2270 & 570 & 308 & 143 & 98 & 42 & 7 
& 1.25 & 6.25  \\
A:$\tilde{q}\tilde{g}$ & 0.28 & 10k & 1440 & 1222 & 684 & 523 & 519 & 387 
& 56 & 0.6 & 3.0 \\
\hline
A:Total  && & &&  & & & &  &1.85 & 9.25\\

\hline
B:$\tilde{g}\tilde{g}$ & 0.06 & 10k & 1987 & 1703 & 1208 & 865 & 789 
& 440 & 136 & 0.4 &2.0 \\
B:$\tilde{q}\tilde{g}$ & 0.018 & 10k & 2702 & 1867 & 1348 & 1153 & 1150 & 889
& 268  & 0.023 & 0.12 \\
\hline
B:Total & & & &&  & & & &  &0.42 &2.12 \\

\hline
C:$\tilde{g}\tilde{g}$ & 0.1 & 10k & 1334 & 909 & 683 & 513 & 466 & 258 
& 68 & 0.2 &1.0 \\
C:$\tilde{q}\tilde{g}$ & 0.57 & 10k & 1057 & 957 & 459 & 330 & 301 & 214 
& 12 & 0.2 & 1.0\\
C:$\tilde{q}\tilde{q}$ & 0.55 & 10k & 882 & 266 & 198 & 139 & 135 & 59 & 2 
& $<1$ & $<1$ \\
\hline
C:Total  && & & & & & & &  &0.4 &2.0 \\

\hline
$t\bar{t}$ & && & &  & & & &  & & \\
\hline
$5-200$ & 143.52 & 100k & 23255 & 1296 & 607 & 78 & 36 & 5 & 0  & 0 & \\
$200-500$ & 16.32 & 50k & 11196 & 2044 & 952 & 382 & 267 & 41 & 2 & 0.27
&1.35  \\
$500-inf$ & 0.16 & 10k & 2030 & 458 & 233 & 152 & 151 & 32 & 5 & 0.02 & 0.1 \\
\hline
Total   & && & & & & & &  & 0.29 & 1.45 \\

\hline
QCD  & & &  && & & &  && & \\
\hline
200-300  & $ 6986$ & 1M & 12413 & 0&0 & 0&0 &0 & 0 &  0 & \\
\hline
300-500  & $ 873$ & 1M  & 1274 & 252 & 13&0&0&0 & 0 &0 & \\
\hline
500-800  &43.1  & 100k &158 & 39&2 & 1&0 &0 & 0 & 0 &  \\
\hline
QCD:Total  & && &  & & && &  & & \\
\hline
\end{tabular}
\caption{
The signal and background events, same as table 3, but for single
lepton final states. The last two columns show the number of events 
multiplied by proper b-tagging efficiency for two luminosity options.}
 \end{center}
\end{table}
\\

{\bf Jets + $\MET$:}
In the simulation for jets+$\MET$ channel, we select at least four jets 
where jets are selected applying cuts as described above. Moreover,  
we apply two new background rejection cuts following the paper of 
ref.~\cite{dipan}. One of them is  
the transverse thrust~\cite{salam} as defined,
\br
{\rm T} = {\rm max}_{n_T}\frac { \sum_i |\vec q_{T,i} . \vec n_T| }
{\sum_i q_{T,i} },
\label{eq:tht}
\er
where the sum runs over all objects in the event,
$\vec q_{T,i}$ is the transverse component of each objects and
$\vec n_T$ is the transverse vector which maximizes this ratio.
Obviously, the events having larger multiplicities will predict 
smaller values of T than the values for the  
events with comparatively lower multiplicities~\cite{dipan}. The other 
variable is related with the ratio of scalar sum of 
transverse of momentum of jets i.e 
the ratio($R_T$) between
the scalar sum of $p_T$ for required lowest number of jets ($n_{j}^{min}$)
in the event and the
scalar sum of $p_T$ of all jets($n_j$) present in the same event,
i.e,
\br
R_{T} = \frac{\sum_1^{n_{j}^{min}} p_T^{j_i}}{H_T}
\label{eq:RT}
\er
with $H_T = \sum_{1}^{n_j} p_{T}^{j_i}$. Evidently, 
behavior of $R_T$ distribution is very much dependent on the number of jets
distributions and their respective hardness in the event . For instance, 
in the case of $n_j \sim n_j^{min}$ which is true for the SM backgrounds, 
$R_T \sim 1$ where as in the case of signal where $n_j >> n_j^{min}$, 
$R_T$ will be smaller than unity. Hence, T and $R_T$ are 
very useful tool to suppress the SM backgrounds without affecting much the
signals rates.

In our simulation for the jets+$\MET$ final states, we apply 
following set of event selection and background suppression cuts,
\br
{\rm Number~of ~ jets} \ge 4;  {\rm T} \le 0.95, {\rm R_T} \le 0.9  
\nonumber \\
\MET \ge 200~GeV; {\rm H_T} \ge 500~{\rm GeV}  \nonumber \\
{\ge \rm 3~b-tagged~jets} {\rm ~with} ~p_T \ge 30~{\rm GeV}, 
|\eta| \le 2.5.  
\er 
In Table 5 we display event yield after each of the cuts for all the 
three  
parameter points 
A,B and C as described in table 1 and also for backgrounds. As before, 
the second and third
column present the production cross sections and the number of events 
simulated for each 
processes respectively. Finally, the number of events normalized by 
luminosities are shown in the last two columns 
with $\ge$3 b-tagged jets in the final states. We see that the 
$\ge$3 b-tags effectively suppresses the background to $<$20\% of the 
signal, so that the viability of the signal is determined by its 
size. Moreover, with a luminosity of 5$\invfb$ we expect 50 signal
events for point A and 15-16 events for points B and C. Therefore, one 
can probe the entire parameter space of this model up to a gluino mass
of $\sim$ 800 GeV.    
% Table 5
\begin{table}
%\begin{center}
\begin{tabular}{|c|c|c|c|c|c|c|c|c|c|}
\hline
Proc&  C.S &N & T &$R_{T}$ &$ \MET $&$H_{T}$  
 & $n_{b}$& 1$\invfb$ & 5$\invfb$    \\
&(pb) &  & $\le 0.95$   & $\le .9$  & $\ge 200  $&$ \ge500$   
& $\ge$3  &  &    \\ 
\hline
A:$\tilde{g}\tilde{g}$ & 5.84 & 10k & 6900 & 3170 & 446 & 301 &  
32 &6.6 & 33.0\\
A:$\tilde{q}\tilde{g}$ & 0.28 & 10k & 8332 & 4356 & 2042 & 2002 & 
315 & 3.4 & 17.0\\
\hline
Total & &  & && &   &  &10 &50\\

\hline
B:$\tilde{g}\tilde{g}$ & 0.06 & 10k & 8708 & 5901 & 2901 & 2584 
& 779 &2.07 & 10.35 \\
B:$\tilde{q}\tilde{g}$ &0.018  & 10k & 9100 & 6263 & 4212 
& 4177 & 1322 & 1.02 & 5.1   \\
\hline
Total & &  & & &  & &  &3.09 & 15.45\\

\hline
C:$\tilde{g}\tilde{g}$ & 0.1 & 10k & 7623 & 3700 & 2888 & 1761 
&404 & 1.13 &5.65 \\
C:$\tilde{q}\tilde{g}$ & 0.57 & 10k & 5933 & 1984 & 1145 
& 1030 & 121 & 1.82 & 9.1 \\
C:$\tilde{q}\tilde{q}$ & 0.55 & 10k & 3326 & 729 & 433 & 385 & 17
&  0.3 & 1.5\\
\hline.
Total & & & & & &  &  & 3.25&16.25 \\

\hline
$t\bar{t}$ & &  & & &  & &  & & \\
\hline
5-200 & 143.5 & 100k & 37162 & 14548 & 21 & 14 & 0 & 0 & 0\\
200-500 & 16.32 & 50k & 27831 & 8559 & 109 & 54 & 3 & 0.48 & 2.4  \\
500-inf & 0.16 & 10k & 1741 & 482 & 68 & 64 & 7 & 0.09 & 0.45\\
\hline
Total & &  & & &  & &  & 0.57 & 2.85 \\
\hline
QCD & &  & &  & &  &  & & \\
\hline
300-500 & 873 & 1M & 194636 & 27532 & 7 & 5 & 4.3 & 0 & 0  \\
\hline
500-800 &43.1 & 100k & 19100 & 2329 & 9 & 9 & 3.8 & 0 & 0 \\
\hline
200-300& 6983 & 1M  & 123 & 14 & 0 & 0& 0 & 0 & 0 \\
\hline
\end{tabular}
\caption{
The signal and background events after, same as table 3, but for jets+$\MET$ 
final states.
The kinematic selection cuts are described in the text. The last two
columns present
the number of events with respective luminosities as shown. 
}
%\end{center}
\end{table}

\section{Summary}
We have analyzed the prospect of probing a non-universal gaugino mass
model of mixed bino-higgsino DM at the current 7 TeV run of LHC. In contrast
to the mSUGRA model this model can provide cosmologically compatible DM
relic density over a large part of the parameter space. In particular 
it provides WMAP compatible DM relic density over two branches 
- (1) $m_{\gl} < m_{\sq}$ and (2)  $m_{\gl} \sim m_{\sq}$ 
-where the DM annihilation occur via gauge boson and Higgs exchanges 
respectively. Moreover, the 1st branch offers promising signals 
for the forthcoming DM detection experiments due to nearly equal 
bino-higgsino components in the DM. On the other hand,
the 2nd branch offers a larger signal at LHC due to the relatively 
light squarks. In contrast to the mSUGRA model, the SUSY mass spectrum of this
model has two distinctive features -(1) an approximate degeneracy of the 
lighter chargino and neutralino masses and (2) an inverted hierarchy of the 
squarks masses. The 1st feature implies that the $\N0_1$ coming from the SUSY
cascade decay carries a relatively large fraction of the energy-momentum, 
resulting in a hard missing transverse energy($\MET$) distribution. 
The 2nd feature implies that the gluino 
decays preferentially to t and b quarks, leading to hard isolated
leptons and multiple b-tags. In particular we have used the $\ge$ 3 b-tags
requirement to effectively suppress the background, so that the viability 
of the signal is determined by the number of signal events.With a 
luminosity of 5$\invfb$, one expects a viable signal in the inclusive 
jets+$\MET$ channel up to $m_{\gl}\sim$800~GeV over both the branches. One
can achieve this in the single lepton + jets+$\MET$ channel with a 
luminosity of 15$\invfb$, which is expected at the end of the current run 
in late 2012. Without the $\ge$3 b-tags, one has to contend with a large
background from $t\bar t$ production. In this case the single lepton
and di-lepton channels offer viable signals up to $m_{\gl} \sim$800~GeV
over the 2nd branch ($m_{\gl} \sim m_{\sq}$) for a luminosity 
of 5$\invfb$, while it goes only upto $m_{\gl}=$400-500~GeV over 
the 1st branch($m_{\gl} < m_{\sq}$).

\section{Acknowledgment}
The authors are thankful to Utpal Chattopadhyay for helping to compute the
SUSY mass spectrum and Debottom Das for helping to make figure. 
DPR acknowledges partial support from the Indian National Science 
Academy through its senior scientist programme.

\end{document}